\documentclass[10pt,twocolumn]{narfront_mod}

\newcommand{\runningtitle}{Impact of point defects on DNA hybridization}

\newcommand{\commentout}[1]{}

\newcommand{\degrees}{${}^{\circ}$}
\newcommand{\degreesC}{${}^\circ$C}

\newcommand{\html}[1]{{ }} 
\newcommand{\htmladdnormallink}[1]{{ }} 

\usepackage{amssymb}
\usepackage{color}
\usepackage{graphicx}
\usepackage{latexsym}
\usepackage{float}
\usepackage{epsfig}
\usepackage{subfigure}

\begin{document}

\begin{frontmatter}



 \title{Hybridization to Surface-bound Oligonucleotide Probes: Influence of Point Defects\newline}


 \author{Thomas Naiser\corauthref{cor}},
 \ead{thomas.naiser@ep1.uni-bayreuth.de}
 \author{Oliver Ehler},
  \author{Timo Mai},
   \author{Wolfgang Michel},

 \and
 \author{Albrecht Ott}
 \corauth[cor]{Corresponding author.}

\address {Experimentalphysik I, Universit\"at  Bayreuth, D-95440 Bayreuth, Germany}
\today

\setlength{\topmargin}{0.5cm}
\setlength{\headheight}{0cm}
\setlength{\headsep}{0cm}
\setlength{\topskip}{0cm}
\setlength{\textheight}{22.7cm}
\setlength{\textwidth}{17.5cm}
\setlength{\oddsidemargin}{-0.5cm}
\setlength{\evensidemargin}{-0.5cm}
\setlength{\columnsep}{0.5cm}
\setlength{\paperwidth}{22cm}
\setlength{\paperwidth}{29.6cm}
\setlength{\hoffset}{0cm}
\setlength{\voffset}{0cm}

\begin{abstract}

Microarray-based genotyping is based on the high discrimination capability of oligonucleotide probes. For detection of  Single Nucleotide Polymorphisms (SNPs) single-base discrimination is required. We investigate how various point-mutations, comprising single base mismatches (MMs), insertions and deletions, affect hybridization of DNA-DNA oligonucleotide duplexes. Employing light-directed \textit{in situ} synthesis we fabricate DNA microarrays with comprehensive sets of cognate point-mutated probes, allowing us to systematically investigate the influence of defect type, position and nearest neighbor effects. Defect position has been identified as the dominating influential factor. 
This positional effect which is almost identical for the different point-mutation types, is biased from the local sequence environment. The impact of the MM type is largely determined by the type of  base pair (either A$\cdot$T or C$\cdot$G) affected by the mismatch. We observe that single base insertions next to like-bases result in considerably larger hybridization signals than insertions next to nonidentical bases.
The latter as well as the distinct position dependence could be explained by a kinetic zipper model in which point defects represent a barrier for the rapid closure of the DNA duplex.

\end{abstract}

\begin{keyword}

hybridization
\sep
defect
\sep
mismatch
\sep
in situ synthesis
\sep
zipper

\end{keyword}

\end{frontmatter}

\bibliographystyle{nar}

running title: \runningtitle




\section*{INTRODUCTION}

DNA microarray technology relies on the highly specific hybridization of complementary nucleic acid strands.
Single-stranded DNA molecules are employed as probes for the detection of complementary target sequences, which are contained within a complex mixture of nucleic acids. Fluorescent labels enable the detection of target molecules captured by their surface-bound counterparts.
High specifity  of probe-target hybridization is required in expression profiling to reduce cross-hybridization from often very similar target sequences. A high discrimination ability is particularly important in genotyping applications, where Single Nucleotide Polymorphisms (SNPs), mutations of single bases, are the subject of interest. SNPs largely determine genetic individuality, but also the individual susceptibility to genetically caused diseases,  and are therefore of great interest not only for genetic research but also for medical diagnostics. \newline
SNPs can be detected by hybridization with short oligonucleotide probes,where already a single mismatching base pair can result in a significant decrease of duplex stability \cite{WALLACE1979}. The hybridization characteristics of mismatched duplexes are governed by many parameters including the position of the defect, type of mismatch, influence of neighboring bases, influence of labeling, secondary structure and  surface effects \cite{Urakawa2003,Lee2004,Karaman2005,Wick2006,Pozhitkov2006}. \newline
A recent study by Pozhitkov et al. \cite{Pozhitkov2006} reveals a poor correlation between predicted and actual hybridization signal (HS) intensities, implying that the thermodynamic properties of oligonucleotide hybridization are by far not yet understood.
Our study is  a comprehensive approach to understand how  point defects affect the  hybridization  of fluorescently labeled DNA oligonucleotide targets to surface-bound probe oligonucleotides.
Applying light-directed \textit{in situ} synthesis  \cite{Fodor1991,McGall1997} with a digital micromirror display (DMD$^{TM}$, Texas Instruments) based maskless synthesis apparatus \cite{Singh-Gasson1999,Gao2001,Nuwaysir2002,Luebke2002,Cerrina2002,Kim03,Baum2003} developed in-house \cite {Naiser2006} we have fabricated DNA chips comprising probe sets with various single base mismatches and other point-mutations with respect to the set of fixed probe sequence motifs employed in this study.
\newline
Following hybridization with labeled targets we determined the relative intensities using fluorescence microscopy.
In order to identify the different factors determining the impact of a defect, we first focused on the strong positional influence \cite{Wick2006,Pozhitkov2006} which can be extracted using a comprehensive set of point-mutated probes. More subtle effects originating from defect type and neighboring bases can be studied after subtraction of the overlying positional dependence. We used  a large set of hybridization signals (normalized and combined from different probe motifs) to analyze characteristics depending on defect type and neighborhood. We further investigate how the positional arrangement of multiple point-defects affects hybridization behaviour.
\section*{MATERIALS AND METHODS}
\subsection*{Reagents}
Hybridization buffer: $5\times$SSPE with 0.1\% SDS, pH 7.4. Cy3-labeled target oligonucleotides - see Table \ref{tab:TargetTable} - were synthesized by MWG Biotech (Ebersberg, Germany) and by IBA Nucleic Acids Synthesis (G\"ottingen, Germany). Oligonucleotide target concentration in the experiments was 1 nM.
\newline
\begin{table}
 \centering
\caption{Fluorescently labeled oligonucleotide targets used in this study}
\begin {tabular} {lllc}

\hline
Name & Target sequence (5'$\rightarrow$3') & Label & Length (nt) \\ [0.5ex]
\hline
URA & \texttt{ACTACAAACTTAGAGTGCAG...} & 5'-Cy3 &  38 \\
    & \texttt{...CAGAGGGGAGTGGAATTC} &  &  \\
NIE & \texttt{ACTCGCAAGCACCACCCTATCA} & 3'-Cy3  & 22 \\
LBE & \texttt{GTGATGCTTGTATGGAGGAA...} & 3'-Cy3 & 30 \\
    & \texttt{...TACTGCGATT} &  &  \\
PET & \texttt{ACATCAGTGCCTGTGTACTAGGAC} & 3'-Cy3 & 24 \\
BEI & \texttt{ACGGAACTGAAAGCAAAGAC} & 3'-Cy3 & 20 \\
COM & \texttt{AACTCGCTATAATGACCTGGACTG} & 5'-Cy3 & 24 \\
NCO & \texttt{TAGTGGGAGTTGTTAGTGATGTGA} & 3'-Cy3 & 24 \\
\hline
\label{tab:table1}
\end {tabular}

\label{tab:TargetTable}
\end{table}

\subsection*{DNA microarray fabrication}
Oligonucleotide microarrays tailor-made for our experiments were fabricated in-house employing light-directed \textit{in situ} synthesis \cite{Fodor1991,McGall1997}. The DMD based synthesis apparatus \cite{Singh-Gasson1999,Gao2001,Nuwaysir2002,Luebke2002,Cerrina2002,Kim03,Baum2003} is described in Naiser \textit{et al. }\cite {Naiser2006}.
Microarrays were synthesized on a phosphorus dendrimer substrate \cite{LeBerre2003}. For compatibility with phosphoramidite chemistry the substrates aldehyde moieties were reduced (in 3.5 mg/ml aqueous solution of sodium borohydride for 3h) to hydroxyl groups.
A photoreactive monolayer is created by coupling of NPPOC-dT-phosphoramidite. Subsequent light-directed synthesis was performed with NPPOC-phosphoramidites \cite{Hasan97,Beier1999}.
Probe sets for the experiments are derived from 16-25mer probe sequence motifs that are complementary to the set of fluorescently labeled target sequences (Table \ref{tab:table1}) available for this study. On the DNA Chip each probe set (comprising between 64 and 400 features) is  arranged as a closely spaced feature block (Figure \ref{fig:FeatureBlockFig1}) which during the analysis can easily be imaged as a whole.
Compact arrangement reduces position-dependent systematic errors that can originate from gradients introduced during synthesis and/or hybridization (see below).
DNA chips produced for this study typically comprise about 2000-3000 features. A relatively large feature size of 21 $\mu$m (6$\times$6 DMD pixels) is used for improved quantitativity.
\subsection*{Real-time monitoring of DNA hybridization}
Hybridization of fluorescently labeled targets to surface-bound probes is carried out in a temperature-controlled hybridization chamber. The chip, synthesized on a 20 mm diameter cover glass (glue-fixed onto a stainless steel support), constitutes a window into the chamber. The chamber volume of 150 $\mu$l is formed by a cutout in a 1.5 mm sheet of PDMS silicone rubber.
Temperature is controlled with a foil heater attached to a stainless steel plate composing the backside of the hybridization chamber.
Relative intensities within the probe sets are largely independent of the hybridization time, chosen to be 10 minutes, typically. Probe sequence motifs with  small hybridization affinities are hybridized for up to 30 minutes to achieve a sufficiently large HS/background ratio.
Hybridization temperature for 16mer probes was typically 30\degreesC. An increased hybridization temperature of 40\degrees C has been applied for probes complementary to the target URA. At 30\degrees C these, due to their large hybridization affinity, hybridize with reduced defect discrimination.
Probes with a length of 20 and more bases are hybridized at 40\degreesC.
Hybridization is monitored in real-time on an Olympus IX81 fluorescence microscope. Thereby the microarray remains in the hybridization solution. A 10$\times$0.4NA UPlanApo objective provides a sufficiently large field of view. An electron multiplying CCD camera (Hamamatsu EM-CCD 9102) with a 1000$\times$1000 pixel resolution is used for image acquisition. During image acquisition flat field correction is performed to compensate for intensity inhomogeneities in fluorescence excitation.

\subsection*{Image analysis}
Image analysis software developed in-house reads the intensities of hundreds of features simultaneously. By selecting corner-points a readout-grid is placed manually onto the image of the hybridized chip. Field distortions are compensated by the trapezoidal arrangement of the grid points (which is determined by the placement of the four corner points).
\begin{figure}[htbp]
  \centering
   \includegraphics[width=8.4cm]{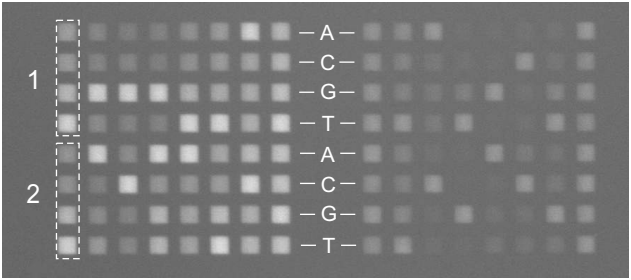}
  \caption{
Fluorescence micrograph of hybridized features in the 16mer mismatch experiment. The shading-corrected image shows two feature blocks corresponding to two different 16mer probe sequence motifs (3'-TTGAGCGATATTACTG-5' - to the left,    3'-TATTACTGGACCTGAC-5' - to the right) both hybridizing with the fluorescently labeled  target sequence COM (5'-Cy3-AACTCGCTATAATGACCTGGACTG-3'). Feature size is 21 $\mu$m. Each feature block comprises all single base mismatches of the particular probe sequence. Groups of four  features (as indicated by the marked groups 1 and 2) correspond to each one the 16 possible mismatch base positions. As indicated by the letters between the feature blocks the uppermost row of features in each group corresponds to an A base at the corresponding base position, followed by probes with C, G and T (see also Figure \ref{fig:ExpSketchFig2}A). The brightest feature within each group corresponds to the perfect matching probe. Nonhybridized targets in the hybridization solution contribute to the background intensity between the features.
Mismatch intensity profiles for the probe sequence motif 3'-TATTACTGGACCTGAC-5'  are shown in Figure \ref{fig:COM5MMProfileFig3}.
  }
  \label{fig:FeatureBlockFig1}
\end{figure}
At the grid points  pixel intensities are integrated over an area typically 10$\times$10 pixels wide, thus excluding the relatively inhomogeneous edge regions of the  21 $\mu$m sized features (Figure \ref{fig:FeatureBlockFig1}). 
The feature raw intensities $I_{F}$ need to be corrected for the solution background intensity $I_{sol}$, which is superimposed on the surface hybridization intensity. To determine the local solution background the read-out grid points are slightly shifted for placement on the gaps between the features. The pure surface hybridization signal $I_{hyb,surf}$ is obtained by subtraction of the solution background from the measured feature intensity. Subsequent normalization (equation \ref{eq:normalization}) through division by the background intensity reduces intensity variations due to lamp flicker.
\begin{equation}
  I_{hyb,surf}=\frac{I_{F}-I_{sol}}{I_{sol}}
  \label{eq:normalization}
\end{equation}
\subsection*{Experiments}
\begin{figure}[htbp]
  \centering
    \includegraphics[width=8.4cm]{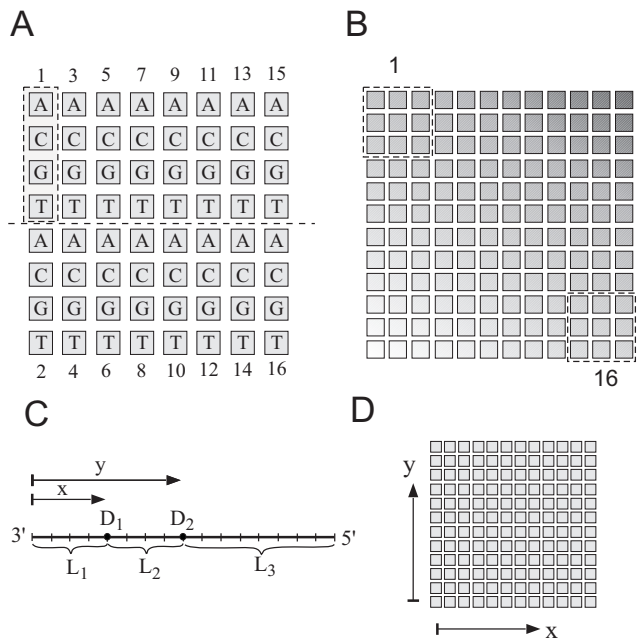}
  \caption{
Microarray feature arrangements \textbf{(A)} for the single base mismatch  experiment (compare with Figure \ref{fig:FeatureBlockFig1}) and \textbf{(B)} for the comparison of various defect types. The gradient indicated in \textbf{(B)} demonstrates that the erroneous variation within the closely spaced feature set belonging to a particular defect position (as depicted with dashed boxes for positions 1 and 16) is significantly smaller than features located far apart. \textbf{(C)} In the two-defect-experiment defects  $D_{1}$ and $D_{2}$ at varying positions x and y divide the sequence into three subsequences of length $L_{i}$. \textbf{(D)} The probe set comprising all configurations of the two defects is arranged in matrix form. Compare with the results in Figure \ref{fig:TwoDelHSFig11}.
  }
  \label{fig:ExpSketchFig2}
\end{figure}

The flexibility of the \textit{in situ} synthesis  and the excellent spot homogeneity simplifies a comprehensive comparative analysis with the capability to detect subtle differences of the probe affinities.
The experiments mainly differ in selection and spatial arrangement of the probe sequences.
Particular experiments focus on the extraction of the positional dependence, the comparison of different defect types and on the identification of further influential parameters.
\newline
Spatial variations of the photodeprotection intensity and optical aberrations affecting the imaging contrast can result in gradients (as indicated in Figure \ref{fig:ExpSketchFig2}B) of the probe DNA quality (due to a varying number of synthesis errors). Thus, for a reliable determination of subtle differences in hybridization affinities, probes to be compared directly should be closely spaced on the microarray.
In the following we describe the design of the individual experiments:
\newline
a) Single base mismatch study
\newline
To investigate the positional dependence of single base mismatches and the impact of the mismatch type, we designed microarrays containing comprehensive sets of MM probes derived from a series of 25 16mer probe sequence motifs. Position and type of the mismatch base pair were systematically varied, allowing us later to distinguish between the dominating positional dependence and other influential factors.
The features are arranged in groups of four, corresponding to the four possible substituent bases (A, C, G and T) at a particular base position. A group comprises three mismatch probes plus one perfect match probe (PM) used for control. Sixteen of these feature groups (one for each base position) are arranged in a square feature block comprising in total 64 features (Figures \ref{fig:FeatureBlockFig1} and \ref{fig:ExpSketchFig2}A).
\newline
b) Single base bulges
\newline
Single base insertions and deletions, due to an extra unpaired base result in bulged duplexes with reduced stability. A comprehensive study on the impact of single base insertions was performed. The experiment comprised about 1000 single base insertion probes (insertion base type and position systematically varied) derived from twelve 20 to 25mer probe sequence motifs.
\newline
c) Comparison of different point-defects
\newline
An experiment allowing for a direct comparison of  PM, MM, single base insertion and deletion probes has been performed. Probe sets were derived from 16mer probe sequence motifs, complementary to the targets in Table \ref{tab:table1}. For each of the 16 possible defect positions a set of 9 probes (comprising four single base insertions, one base deletion, three MMs and one PM probe) has been created. To avoid that a regular arrangement of the probe features could possibly affect the measurement (e.g. by introducing a bias due to increased target depletion near a PM probe), the sets of nine probes were randomly arranged in 3$\times$3 matrices.
\newline
d) Influence of the positional distribution of multiple defects
\newline
To investigate how the distribution of two or more defects  affects duplex stability we followed two approaches:
We designed a probe set comprising all two-deletion mutations of a 20mer probe sequence motif: Deletions $D_{1}$ and $D_{2}$ (as shown in Figure \ref{fig:ExpSketchFig2}C) were introduced at positions x and y. The positions of the defects were independently varied from  base positions 1 to 20, resulting in a 20$\times$20 matrix (Figures \ref{fig:ExpSketchFig2}D and \ref{fig:TwoDelHSFig11}) of 400 probes comprising all two-deletion probes in duplicate (plus 20 single deletion probes - for x and y coinciding).
To extract the pure influence of the defect configuration on the HS, averaging was performed over a set of nine different 20mer motifs to eliminate sequence specific bias.
\newline
For larger numbers of defects we applied a statistical approach. Based on a 20mer probe motif we created sets of randomly mutated probes containing a fixed number of one to five base deletions at random positions.
\section*{RESULTS}
\subsection*{Influence of the point defect position}
\begin{figure}[htbp]
  \centering
\includegraphics[width=8.4cm]{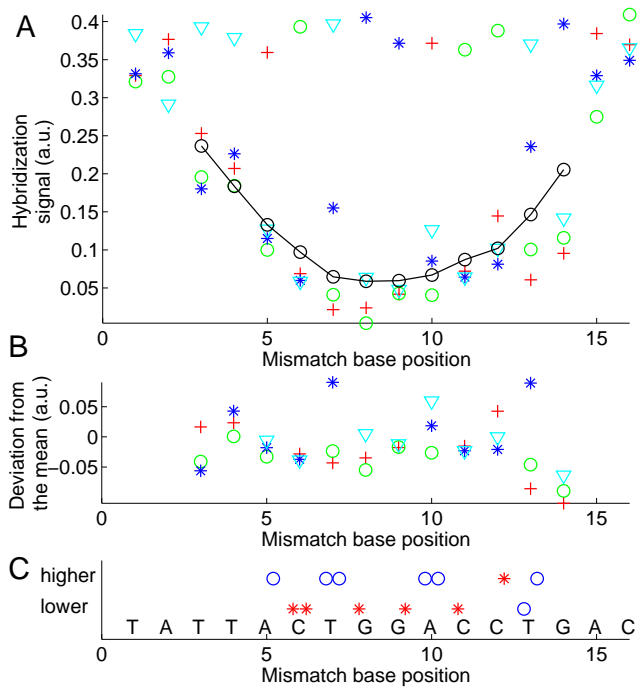}
  \caption{
Mismatch intensity profile \textbf{(A)} (HS vs. defect base position) obtained from the hybridization signals of the feature block shown in the right part of Figure \ref{fig:FeatureBlockFig1}. The probe sequence motif 3'-TATTACTGGACCTGAC-5' is complementary to the target oligonucleotide COM. Markers depict the substituent base type (A red crosses; C green circles; G blue stars; T cyan triangles). The black line indicates the mean profile (moving average of all mismatch HS over positions  p-2 to p+2). PM probes, included as control to detect erroneous bias, have the largest HS (at a level of about 0.38 a.u.). The variation of the PM probe intensities also provides an estimate for the error of the measurement. Errors between distant features, due to gradient effects, are expected to be larger than errors between the compactly arranged features corresponding to a particular defect position.
\textbf{(B)} Deviation profile. The strong position dependent component of the HS is eliminated by subtraction of the mean profile.
\textbf{(C)}
Comparison of mean mismatch HS (average of the three mismatch HS at a particular defect position) at the sites of C$\cdot$G base pairs to mean MM hybridization signals at the site of adjacent A$\cdot$T base pairs. A marker (red star: A$\cdot$T; blue circle C$\cdot$G)is set in the upper row if the HS of the mismatches at the corresponding site is higher than that  at the adjacent site; otherwise a marker is set in the lower row. We noticed that mismatches substituting a C$\cdot$G base pair usually have systematically lower HS than mismatches substituting a neighboring A$\cdot$T base pair.
  }
  \label{fig:COM5MMProfileFig3}
\end{figure}

\begin{figure*}[h!!tbp]
  \centering
    \includegraphics[width=14cm]{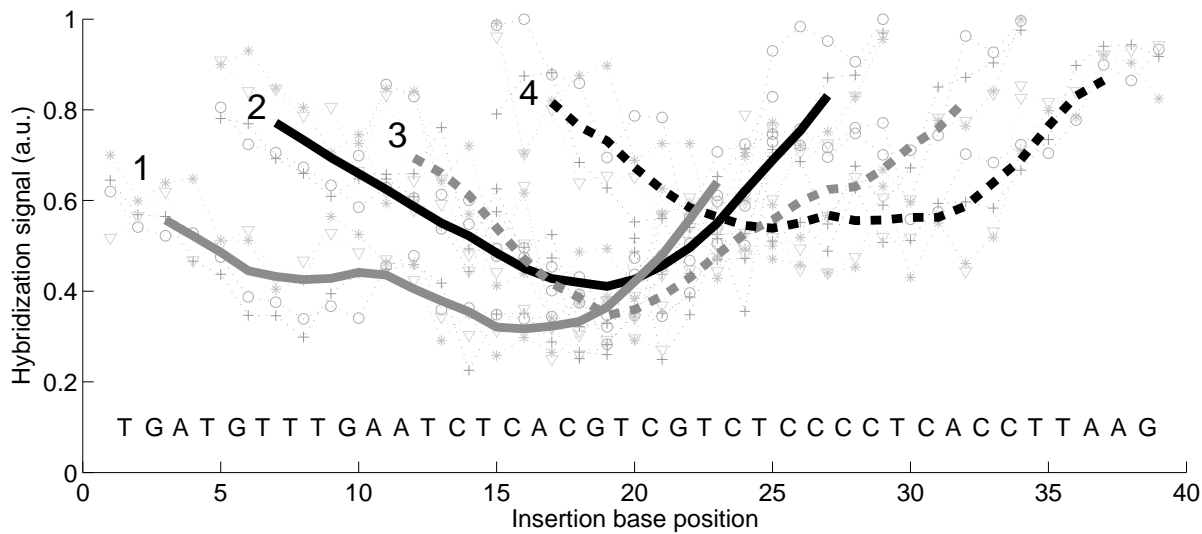}
  \caption{
The impact of defects is also affected by the local sequence environment. Normalized single base insertion profiles (hybridization signal plotted versus the insertion base position) of four 25mer probe sequence motifs complementary to the same target URA. The probe motifs 1 to 4 hybridize at different sections of the target oligonucleotide. Mean profiles (thick lines) were obtained  from the moving average of the particular insertion profiles (particular HS are shown as faint symbols - profile 4 is shown in detail in Figure \ref{fig:InsProfileFig6}A). The mean profiles 1 to 3 have a distinct minimum between base positions 15 to 20. The stabilizing CG-rich region following after base position 20 results in increased HS in profile 4.
  }
  \label{fig:LocalSeqEnvFig4} 
\end{figure*}

From the profile plots (plots of the normalized hybridization signal vs. defect position - cf. Figures \ref{fig:COM5MMProfileFig3} and \ref{fig:InsProfileFig6}) we see that the most influential  parameter determining hybridization intensity is the position of the defect. We observe that the intensities of the particular mismatch probes (Figure \ref{fig:COM5MMProfileFig3}) are lined-up along the mean profile, which can be roughly approximated by a second degree polynome.
\newline
Defects in the middle of the probes are most destabilizing. For 16 mer duplexes a single MM in the center typically results in 0-40\% of the PM hybridization signal. Whereas defects near the edges only have a small effect.
Nonpositional factors (e.g. mismatch type and nearest neighbor effects) have significantly smaller influence on the HS than the defect position. Variation of the intensity profiles due to mismatch type and other nonpositional factors is usually less than 20\% of the PM intensity level.
The discrimination between PM and point-mutated probes depends on the stability of the particular probe sequence. Shorter 16mer sequences (Figures \ref{fig:COM5MMProfileFig3}A and \ref{fig:ProfileCompFig9}) are more discriminative than the more stable 25mer probes (shown in Figure \ref{fig:InsProfileFig6}). Increased stability can also be achieved with a high CG content of the probes.  Probe sequence motif 4 in Figure \ref{fig:LocalSeqEnvFig4} demonstrates that a high CG content results in increased perfect match HS (due to normalization not noticeable here), albeit with reduced discrimination.
The positional influence observed in the mean profiles is largely determined by the defect-to-end distance, but it is furthermore superimposed by a sequence dependent contribution.
The variation of the shapes of the mean insertion profiles in Figure \ref{fig:LocalSeqEnvFig4} indicates that the impact of a defect is affected by the stability of the local sequence environment.

We observe the same strong positional dependence for single base insertions (Figure \ref{fig:InsProfileFig6}) and deletions (Figure \ref{fig:ProfileCompFig9}).
For a systematic study of other influential parameters (e.g. defect type and defect neighborhood) the dominating positional influence has to be eliminated.
Design (selection and arrangement of probes) and analysis of our experiments therefore focus on the separation of the different influential factors.

\subsection*{Effect of the mismatch type}

\begin{figure}[h!!tbp]
  \centering
    \includegraphics[width=8.4cm]{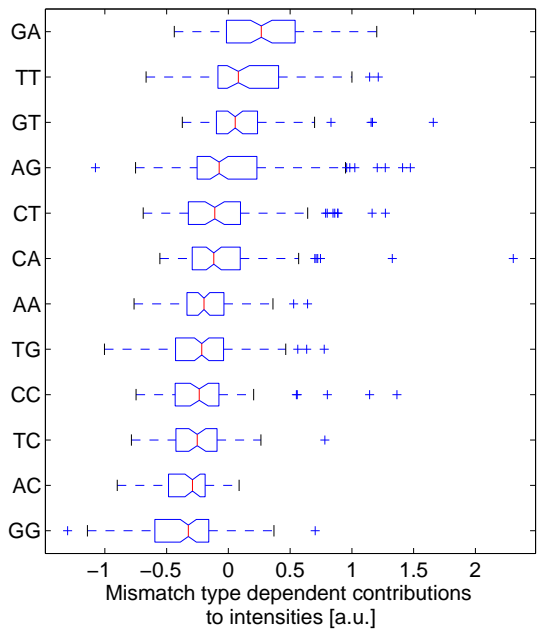}
  \caption{
Boxplot representation of the hybridization signal distributions for the individual mismatch types, arranged according to the median values (depicted by the vertical line at the notch). Boxes indicate the interquartile range (from the 25th to 75th percentile) containing 50\% of the data. Whiskers extend to a maximum value of 1.5 times the interquartile range from the boxes ends. Values beyond are classified as outliers. If the notches of two boxes do not overlap the medians values differ significantly  with a 95 percent confidence.
The mismatch types with the lowest hybridization signals are those (T$\cdot$G, C$\cdot$C, T$\cdot$C, A$\cdot$C, G$\cdot$G) where C$\cdot$G base pairs are affected by the mismatch defect. The only exception is A$\cdot$G. The positive tails of this and other distributions seem to originate from  stabilizing C$\cdot$G base pairs next to the defect.
}
  \label{fig:MMTypeBoxPlotFig5}
\end{figure}

For the analysis of MM type and nearest-neighbor influences the positional influence was eliminated by subtraction of the mean profile. The resulting 'deviation profile' (comprising nonpositional influences) is shown in Figure \ref{fig:COM5MMProfileFig3}B.
\newline
In the following we use the notation of the mismatch base pair X$\cdot$Y consisting of the mismatching base X in the probe sequence and the base Y in the target sequence.
A pairwise comparison of the averaged mismatch HS (mean value of the hybridization signals of the three MM probes corresponding to a particular defect position - see Figure \ref{fig:COM5MMProfileFig3}B) at directly adjacent defect positions  reveals that mismatches affecting  C$\cdot$G base pairs decrease the intensities significantly more than mismatches affecting a neighboring A$\cdot$T base pair (Figure \ref{fig:COM5MMProfileFig3}C).
The magnitude of this effect is typically at 5 to 10\% of the perfect match HS.
To investigate how the various MM-types X$\cdot$Y affect duplex stability we use data from 25 different probe sequence motifs.
The PM hybridization signals of the different 16mer probe sequence motifs display a strong variation (up to a factor of 20), and are therefore not directly comparable. Since the relative intensities (of the various MM probes) within the probe sets are largely unaffected by this variation, we can normalize the 'deviation profiles' by division with their standard deviation. The resulting database comprising normalized hybridization signals (with the positional influence eliminated) from about 1000 different single MM probe sequences, is categorized according to mismatch type.
The boxplot representation   of this data (Figure \ref{fig:MMTypeBoxPlotFig5}) shows that MM-types affecting C$\cdot$G base pairs  (A$\cdot$C, C$\cdot$C, T$\cdot$C and A$\cdot$G, G$\cdot$G, T$\cdot$G) systematically have lower median hybridization signal values than those MM-types affecting  A$\cdot$T base pairs (A$\cdot$A, C$\cdot$A, G$\cdot$A and C$\cdot$T, G$\cdot$T, T$\cdot$T).\newline
A mismatch base substituting C or G is statistically more destabilizing than a MM base replacing  A or T, indicating that C$\cdot$G base pairs are more crucial for duplex stabilization than A$\cdot$T base pairs.
This leads to obvious differences between the distributions A$\cdot$C/C$\cdot$A, A$\cdot$G/G$\cdot$A, T$\cdot$C/C$\cdot$T and T$\cdot$G/G$\cdot$T. Although the mismatch types  X$\cdot$Y and Y$\cdot$X may be thought to be equivalent (because the bases involved are the same), they result in different PM/MM hybridization signal ratios, depending on  the type of base pair (A$\cdot$T or C$\cdot$G) affected by the  mismatch. For example, the impact of the MM  A$\cdot$C affecting  an A$\cdot$T base pair is (on average) smaller than the impact of the MM C$\cdot$A affecting a C$\cdot$G base pair. Thus the ratio of MM to PM hybridization signals, i.e. the relative stability of the MM duplex in comparison to the corresponding PM duplex is larger for the mismatch A$\cdot$C than for the mismatch C$\cdot$A:
	\[
	\frac{HS_{MM=A\cdot C}}{HS_{PM=A\cdot T}} > \frac{HS_{MM=C\cdot A}}{HS_{PM=C\cdot G}}
\]
\subsection*{Nearest neighbor influence}
The distribution of HS of the mismatch type A$\cdot$G  which has a distinct maximum at negative values (i.e. HS values below the average, cf. Figure \ref{fig:MMTypeBoxPlotFig5}) extends also far towards the positive side, indicating many occurrences of increased HS values. This tail of increased hybridization signals largely corresponds to A$\cdot$G mismatches with nearest neighbors (right and left to the mismatch base)  CC, CT and TC.
For a systematic study the mismatch HS data is now categorized not only for the mismatch type (as discussed above), but also for the nearest neighbor bases (i.e. the bases next to the mismatching base). There are 16 neighborhood categories for each of the 12 mismatch  types. Since our database is too small for a detailed analysis of all neighborhood categories, we simplified the neighborhood classification to only three neighborhood categories: A$\cdot$T  base pairs only, C$\cdot$G base pairs only and mixed neighborhoods of A$\cdot$T and G$\cdot$C base pairs. For 9 of the 12 mismatch types the HS are significantly increased if the base pairs adjacent to the defect comprise at least one C$\cdot$G base pair. Exceptions are the mismatches G$\cdot$T, C$\cdot$T and A$\cdot$A. We also noticed a tendency that mismatches with two C$\cdot$G-neighbors have larger HS than such with only one C$\cdot$G-neighbor, but, to be significant, this result needs to be corroborated with more data.
\newline
The largest neighborhood-related variations are observed for X$\cdot$G - mismatches (with X=A,G or T), whereas X$\cdot$T-mismatches consistently show the smallest neighborhood-related variations.
\subsection*{Single base bulge defects}
\begin{figure}[htbp]
  \centering
   \includegraphics[width=8.4cm]{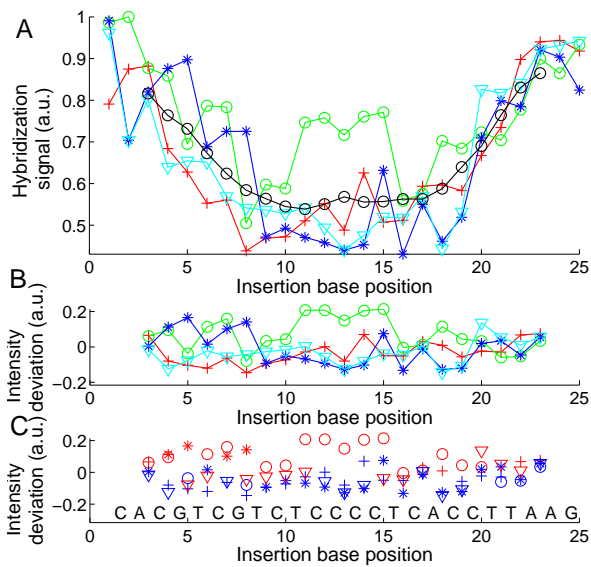}
  \caption{
\textbf{(A)} Single base insertion profile (hybridization signal plotted versus insertion base position) of the probe sequence motif 3'-CACGTCGTCTCCCCTCACCTTAAG-5' (complementary to the target URA). Symbols correspond to insertion bases (A red crosses; C green circles; G blue stars; T cyan triangles). The  mean profile (black line), obtained from the moving average (including all 4 insertion types) over  positions p-2 to p+2 shows the common positional dependence. Insertions to the left and to the right of an identical base (Group II bulges - see text) result in identical probe sequences.
\textbf{(B)} and \textbf{(C)} Deviation profiles. Positional influence is elimated by subtraction of the mean profile. Elevated intensities are observed for Group II bulges (e.g. C insertions at positions 11 to 15, 6 to 7 and  18 to 20 or G insertions at positions  4 to 5 and 7 to 8). A very distinct increase of the HS is observed for C insertions into the subsequence TCCCCT. Group II bulges (red  markers) have significantly higher intensities compared to Group I  bulges (blue markers).
  }
  \label{fig:InsProfileFig6}
\end{figure}

\begin{figure}[htbp]
  \centering
    \includegraphics[width=8.4cm]{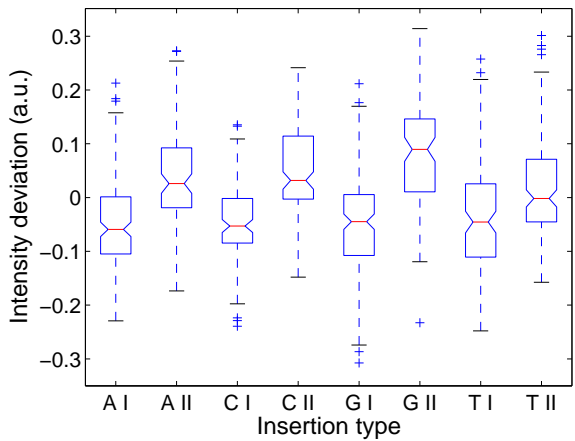}
  \caption{
Boxplots show the hybridization signal deviations (from the mean profile) for the different insertion base types, which are differentiated according to affiliation to bulge Group I/II. The statistical analysis includes about 1000 normalized hybridization signals from 12 different 20 to 25mer probe sequence motifs. 
}
  \label{fig:InsBoxPlotFig7}
\end{figure}

Single base insertions and deletions result in bulged duplexes with reduced stability. In duplexes with single base insertion probes the bulged base is located on the surface-bound probe strand, whereas for single base deletion probes the bulged base is located on the target strand.
The positional dependence of the insertion intensity profiles (Figure \ref{fig:InsProfileFig6}A) is very similar to the mismatch intensity profile in Figure \ref{fig:COM5MMProfileFig3}, though the individual insertion profiles (for example the profile of C-insertions - green circles in Figure \ref{fig:InsProfileFig6}) show large deviations from the mean (moving average) profile. Hybridization signals  are significantly increased over two or more consecutive defect positions. These defects, corresponding to base insertions next to identical bases, result in relatively stable duplexes. This effect was previously described by Ke \textit{et al.} \cite{Ke1995}.
In the notation of Zhu \textit{et al.} \cite{Zhu1999} bulged bases without an identical neighboring base are defined as  Group I bulges, whereas bulges with at least one identical neighboring base are referred to as Group II bulges.
We found the stabilizing effect of Group II bulges (in respect to Group I bulges) to be surprisingly large. For Group II bulges located near the middle of 16mer probes hybridization signals approaching that of the corresponding perfect match probe are not  unusual.
Figure \ref{fig:InsProfileFig6}C demonstrates the systematically increased stability of Group II bulges.
A statistical analysis with a larger dataset (Figure \ref{fig:InsBoxPlotFig7}) comprising HS data from 1000 different probes indicates the general validity of the result. The median HS values of Group I insertions do not significantly differ with the base inserted.
Compared to these results, in a similar experiment with 16mer probes (cf. Figure \ref{fig:InsBoxPlotFig7}) the HS of Group I C insertions were significantly increased, whereas HS of Group I A insertions were reduced to that of mismatches.
The lower hybridization temperature of 30\degrees C  (compared to 40\degrees C for the longer 20-25mers) could be responsible for the difference \cite{Karaman2005}.
Largest differences in the 25mer experiment (corresponding to about 20\% of  the PM hybridization signal) between Group I and Group II bulges are observed for G-insertions , whereas a significantly smaller difference (of only 5\%) is found for T-insertions.
At the ends of the duplex the average difference between Group I/Group II HS is reduced to about 50\% of the value that is observed for defects in the middle of the duplex.
In case of single base deletions (e.g. in Figure \ref{fig:ProfileCompFig9}, orange dashed line) HS for Group II bulges are increased, albeit the difference between Group I and Group II bulges unexpectedly is distinctly smaller than for insertion defects. Interestingly, the deletion profiles are well confined within the HS range spanned by the various mismatch defects.

\begin{figure}[htbp]
  \centering
    \includegraphics[width=8.4cm]{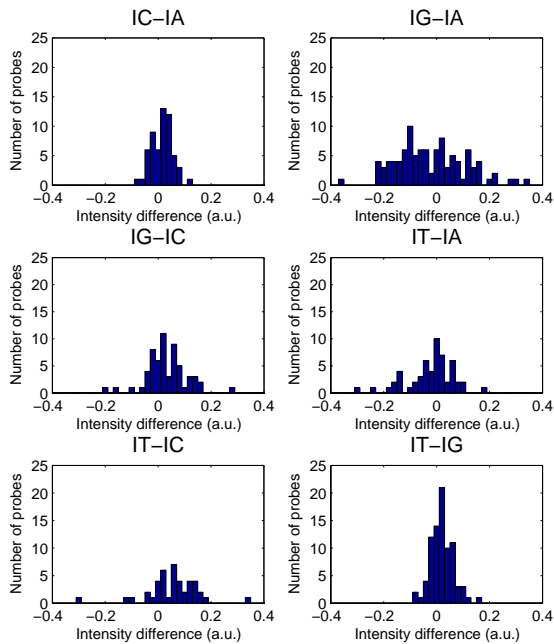}
  \caption{
Histograms of hybridization signal differences IX-IY (X and Y denote the different insertion bases in otherwise identical probe sequences) reveal correlations between the HS of different insertion types. To exclude the impact of systematically increased intensities of Group II insertions only Group I insertions are regarded here. Between T- and G-insertions (and between C- and A-insertions) a correlation, as indicated by a narrow distribution with a pronounced peak near zero, is observed. The broad distribution of HS differences between G and A insertions doesn't show a distinct peak, indicating that there is no correlation but rather an anticorrelation for insertions of A and G.
 }
  \label{fig:InsHistPlotFig8}
\end{figure}
Systematically increased HS are not restricted to Group II base bulges. For G-insertions next to a T base (e.g. in Figure \ref{fig:InsProfileFig6} at base position 15) HS are often  significantly increased, similar as with Group II bulges.
\newline
We have further investigated the degree of correlation between probes with different insertion bases (Figure \ref{fig:InsHistPlotFig8}). A distinct correlation is seen between T- and G-insertions, and, though less distinct, between  A- and C-insertions. In contrast to that our results indicate an anticorrelation between G- and A-insertions.
\subsection*{Comparison of single base insertion, deletion and mismatch defects}
\begin{figure}[htbp]
  \centering
    \includegraphics[width=8.4cm]{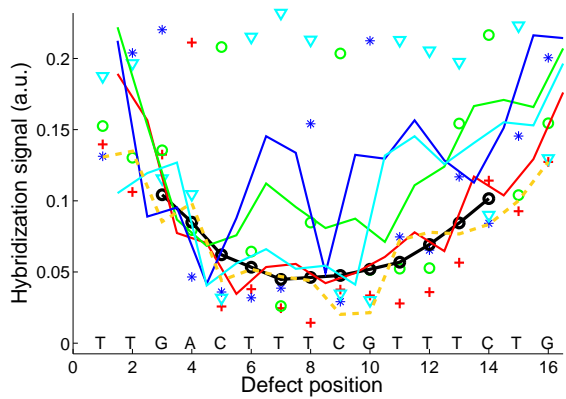}
  \caption{
Comparison of  single base mismatches, insertions and deletions. The 16mer probe sequence motif 3'-TTGACTTTCGTTTCTG-5' is complementary to the target BEI. Normalized mismatch probe intensities with substituent bases A (red crosses), C (green circles), G (blue stars), T (cyan triangles), running average of mismatch intensities (black line); normalized single base insertion probes (solid lines) with insertion bases A (red), C (green), G (blue), T (cyan). Hybridization signals of single base deletions (orange dashed line) are comparable to that of mismatches at the same position. Increased HS of certain insertion defects are due to positional degeneracy (see discussion) of base bulges.
  }
  \label{fig:ProfileCompFig9}
\end{figure}

\begin{figure}[htbp]
  \centering \includegraphics[width=7.5cm]{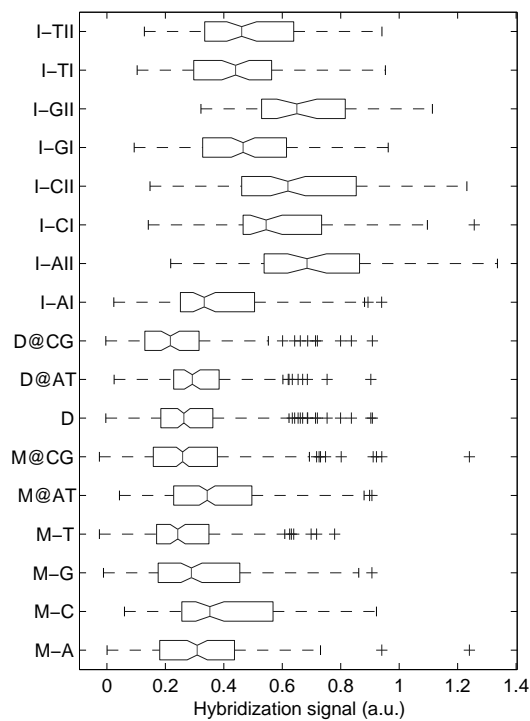}
  \caption{
Comparison of the hybridization signals of different point mutation types. To minimize positional influence the statistics include only defect positions 5 to 12, located in the center of the 16mer probes. The 1200 probe sequences were derived from 17 probe sequence motifs. HS are normalized in respect to the perfect match HS.
Defect categories:
mismatch M-X (X: substituent base);  mismatches at A$\cdot$T and C$\cdot$G sites M@AT, M@CG; single base deletion D; deletions at A$\cdot$T and C$\cdot$G sites D@AT, D@CG; single base insertion I-XI/II (X: insertion base, I/II: Group I/Group II base bulge).
Hybridization signals from insertion probes (about 50\% of the PM hybridization signal for Group I; 65\% for Group II - median values) are significantly higher than that of MM probes (at about 30\%). Mismatches at A$\cdot$T sites result in about 25\% larger HS than MMs at C$\cdot$G sites. Deletion probes have a median hybridization signal that is slightly lower than the median MM hybridization signal.  Group I base bulges with the exception of I-AI (33\%) have HS of about 50\% of the PM hybridization signal. Hybridization signals of Group II base bulges are significantly higher (about 100\% for A insertions, and only 5\% for T insertions) than that of the corresponding Group I bulges.
}
  \label{fig:HSCompBoxPlotFig10}
\end{figure}

A direct comparison of HS of different point defects (Figure \ref{fig:ProfileCompFig9}) reveals that the positional influence is largely independent of the defect type.
\newline
Single base insertion probes provide distinctly larger hybridization signals than MM probes (Figure \ref{fig:HSCompBoxPlotFig10}). This is partly due to the reduced number of binding base pairs in the mismatched duplexes, but also due to the significantly increased HS of Group II insertions.
Hybridization signals of MMs inserted into C$\cdot$G base pairs are about 25\% smaller than those of MMs inserted into A$\cdot$T base pairs.
Similarly, single base deletions affecting C$\cdot$G base pairs result in about 30\% smaller HS  than deletions affecting  A$\cdot$T base pairs. This can also be observed in the deletion profile in Figure \ref{fig:ProfileCompFig9} (orange dashed line), where HS from deletions in C$\cdot$G/A$\cdot$T base pairs result in lower/higher HS. The fact, that the same effect has been observed for MMs suggests, that in both cases the defect-type-dependent impact on the hybridization signal is largely determined by the type of base pair affected by the point-mutation.

\subsection*{Multiple defects - influence of defect distance}
\begin{figure}[h!!!!]
  \centering
    \includegraphics[width=8.4cm]{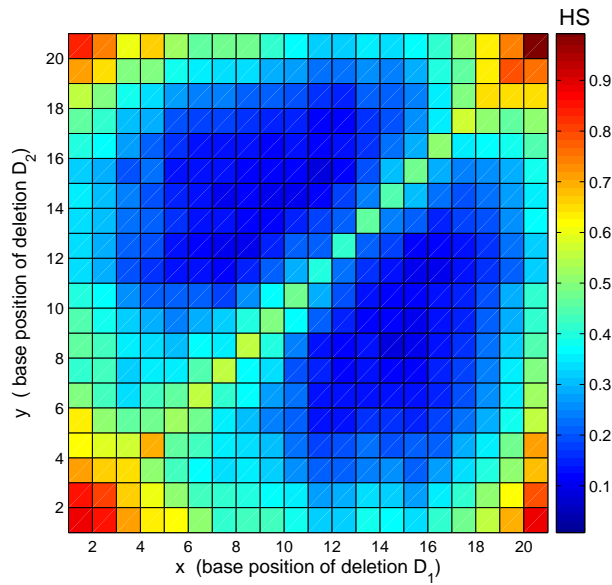}
  \caption{
Hybridization signals  of 20mer probes (normalized in respect to the maximum HS) with two single base deletion defects $D_{1}$ and $D_{2}$ at varying positions x and y (compare to \ref{fig:ExpSketchFig2}D). Averaging over data sets obtained from 9 different probe sequence motifs has been performed to eliminate nonpositional contributions (e.g. differences resulting from deletions affecting  either A$\cdot$T or C$\cdot$G base pairs) from the HS. The resulting data set shows the influence of the defect distribution on the hybridization signal.
Defects at the probe 3'-end (base position 1) affect the HS slightly less than defects at the 5'-end.
  }
  \label{fig:TwoDelHSFig11}
\end{figure}

\begin{figure*}[htbp]
\centering
\includegraphics[width=16.0cm]{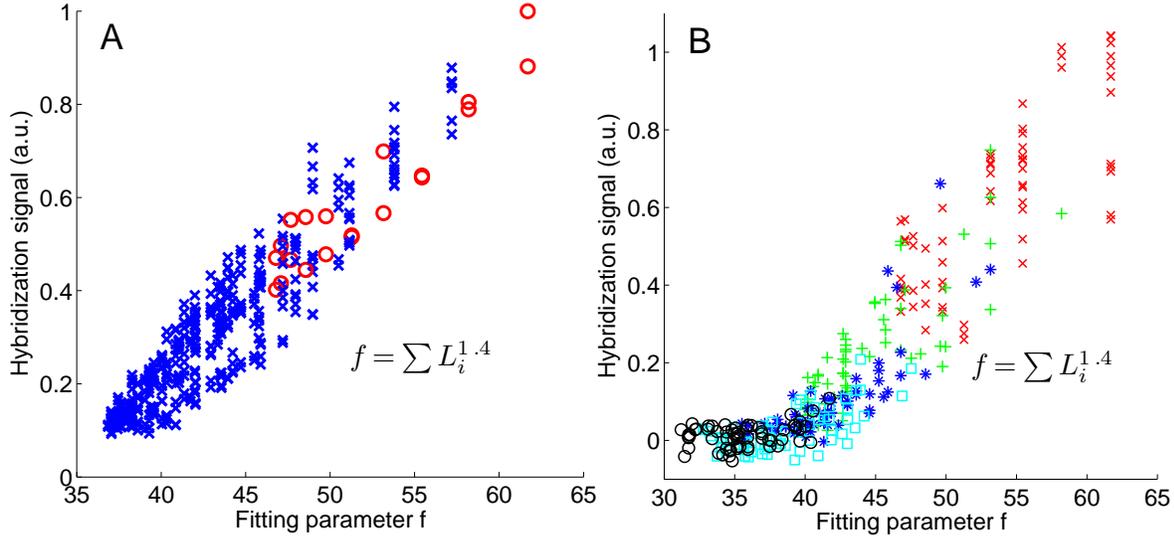}
\caption{
Fitting of the hybridization signals of multiple-defect probes. \textbf{(A)} Hybridization signals obtained from the two-deletion experiment are plotted versus the fitting parameter $f=\sum L_i^{1.4}$ ($L_i$: length of defect-free subsequences - see Figure \ref{fig:ExpSketchFig2}C) (two deletions: blue crosses; single deletion - for x and y coinciding: red circles). The fit can further be improved by correction for the 3'-bias observed in Figure \ref{fig:TwoDelHSFig11}. \textbf{(B)} Similar experiment with a varying number of deletions (at randomly chosen positions) in the 20mer probe sequence motif 3'-TAGTCACGGACACATGATCC-5'. Marker types indicate the number of  deletions: 1 red cross; 2 green cross ;3 blue star; 4 cyan square; 5 black circle). Because only data from a single probe sequence motif was available, non-positional (sequence related) contributions  couldn't be eliminated, thus resulting in increased scattering of the HS.
}
\label{fig:MultDelsFig12}
\end{figure*}

In two-deletion experiments (Figures \ref{fig:ExpSketchFig2}C and \ref{fig:TwoDelHSFig11}) we determined the HS of 20mer probes with systematically varied arrangements of two single-base deletions.
The HS is largest when both defects are located close to one or both ends. Lowest hybridization intensities are observed for defect configurations dividing the sequence into three roughly equally long subsequences.
Closely spaced defects (with a distance of less than 4 bases - located near the diagonal of the plot) result in increased hybridization signals approaching that of single base deletions as the distance between the defects is reduced.
Hybridization signals of multi-defect probes can be described  by equation (\ref{eq:fit}).
\begin{equation}
  f=a\cdot \sum L_i^{\nu}+b
  \label{eq:fit}
\end{equation}
Therein $L_i$ denote the lengths of defect-free subsequences - a and b are free parameters. To account for the fact that longer subsequences contribute disproportionately more to the HS than shorter ones, the exponent $\nu$ is introduced, putting a length dependent weighting factor to the particular lengths $L_i$.
In Figure \ref{fig:MultDelsFig12}A  the HS of the two-deletion experiment were plotted versus the parameter f from equation (\ref{eq:fit}).
As shown in Figure \ref{fig:MultDelsFig12}A equation (\ref{eq:fit})  also predicts HS for probes with a larger number of deletions.

\section*{DISCUSSION}
%

\subsection*{Influence of synthesis errors originating from the light-directed \textit{in situ} synthesis process}
Light directed \textit{in situ} synthesis, due to stray light and incomplete coupling reactions, introduces point mutations (MMs, insertions an deletions) in the probe sequences. A large fraction (we estimate about 80\% \cite{Naiser2006}) of the probes contains at least one point defect. Our experiments demonstrate that probes with point defects  do contribute significantly to the HS. The distribution of various point defects contained within a microarray feature results in broadening of the melting transition. This effect has also been observed for mixtures of truncated probe sequences \cite{Jobs2002}. Experiments where we systematically varied the degree of UV deprotection (results not shown) to introduce a  varying number of deletion defects show that duplex stability, melting temperature and hybridization signal intensity are reduced as the number of defects increases.
The statistical nature of the synthesis errors and the large number of different defect types (for a 20mer probe sequence there are about 160 different single base defects) prevents a bias towards a particular type of defect.
Our experimental results demonstrate that despite a large fraction of probes containing random defects, a high discrimination of single base defects can be achieved.
The discrimination between PM and single base defect hybridization signals is determined by duplex length and CG content. With increasing overall duplex stability the relative destabilizing impact of single base defects is reduced.

\subsection*{Influence of the defect position}
We found a dominating influence of the defect position on duplex stability. Hybridization signals are largely determined by a smoothly varying function of defect position.
Defects in the middle of the duplex have significantly smaller HS than defects at the ends of the duplex, which usually result in small or insignificant decreases compared to the PM hybridization signal.
Strong positional influence, has also been reported by other authors:
From optical melting studies (on 7mer RNA/RNA duplexes in solution) Kierzek \textit{et al.} \cite{Kierzek99} report a 0.5 kcal/mol stabilization increment per each base position that the defect is closer to the helix end. A positional influence was observed for U$\cdot$U and A$\cdot$A, whereas G$\cdot$G MM stability was largely unaffected by the position.
Dorris \textit{et al.} \cite{Dorris2003} found a similar positional influence for  2-base MM and 3-base MM probes on CodeLink 3D gel arrays. They observed a strong correlation between solution-phase melting temperatures and microarray hybridization signals of the MM duplexes.
More recently Wick \textit{et al.} \cite{Wick2006} and Pozhitkov \textit{et al.} \cite{Pozhitkov2006} reported a strong influence of the defect position on the HS of surface-bound single base MM probes (fabricated by \textit{in situ} synthesis). They present averaged profiles showing a similar positional influence as observed in our experiments with  individual probe sequence motifs.
In accordance with \cite{Pozhitkov2006} we have identified MM position (relative to the duplex ends) as the strongest influential factor on the HS, when compared to MM type and nearest neighbors.
Position dependence is in contrast to the nearest-neighbor model of DNA duplex thermal stability, where the thermodynamics of internal mismatches is treated as independent of the MM position \cite{SantaLucia2004}.\newline
For single base bulges we observed the same position dependence as for MMs. Also, the magnitudes of the impacts of the  different point-mutations on the HS are very similar (with the exception of Group II bulges, which have significantly higher HS). These coincidences suggest a common origin of the positional influence, which is independent of the defect type.
Focusing on individual probe sequence motifs, we discovered, that the positional influence does not only depend on the defect-to-end distance, but also has a sequence-dependent contribution. This indicates that the impact of a defect is affected by the stability of the local sequence environment (beyond the nearest neighbors). A similar sequence-dependence of the base pair closure probability has been predicted in  zipper models of the DNA \cite{Deutsch2004,Binder2006}. 
\newline
In the following we discuss our results in the context of the opening and closing dynamics of a molecular zipper model of the DNA duplex \cite{Gibbs1959,Zimm1960,Applequist1965,Kittel1969,Craig1971,Ambjornsson2005}.
Due to partial melting propagating from the ends (a high initiation barrier prevents the formation of denaturation bubbles in the interior of the relatively short oligonucleotide duplexes), the base pair closure probability is reduced towards the duplex ends.
\newline 
In a very simple model the influence of defect position on the hybridization signal $HS_{def}(x)$ could be described by equation (\ref{eq:closProb}), in which the PM hybridization signal $HS_{PM}$ is reduced by the position-dependent impact of the defect. The latter is determined by the base pair closure probability $P_{cl}(x)$ and the (defect type dependent) impact $I_{def}$.
\begin{equation}
\label{eq:closProb}
	HS_{def}(x)=HS_{PM}-P_{cl}(x)\ast I_{def}
\end{equation}
In the following we assume that a duplex dissociates once it is completely unzipped.
Then a parabola-like position-dependence, as observed in our experiments, with relatively large base pair opening probabilities $P_{op}(x)=(1-P_{cl}(x))$ near the middle of the duplex, implies an unrealistically large duplex dissociation rate. In fact $P_{op}(x)$ is expected to decay exponentially, approaching a value close to zero in the middle of the duplex. A fast decay of the opening probability (with typically $P_{op}(x)>0.1$ only for the  two or three outermost base pairs) is required for a realistic duplex dissociation rate.
Short duplexes coexist with single strands in a bimolecular equilibrium with only a very small proportion of partially unzipped duplexes present \cite{Zimm1960,Applequist1965,Craig1971}. 
Since the base pair closure probability $P_{op}$ must be very close to one in a bimolecular equilibrium of duplexes and single strands, we must conclude that the above model (equation \ref{eq:closProb}) is too simplistic.
\newline
The results of the two-deletion experiments show that the longer defect-free subsequences contribute disproportionately to the HS signal. For adjacent defects, separated by less than 5 base positions,  hybridization signals are increased. HS of directly adjacent defects (here a merged two-base bulge is expected) are close to that of single base bulges and significantly larger than the smallest HS which are observed for defect distances of 5 to 6 base positions (cf. Figure \ref{fig:TwoDelHSFig11}). Thus the effect of 2 deletions on the HS is not additive but depends strongly on the positional distribution of the defects.
The HS of probes with two deletions is well fitted by equation \ref{eq:fit} (cf. Figure \ref{fig:MultDelsFig12}), basically the weighed sum of the lengths of defect-free subsequences of the duplex. \newline
The connection between the hybridization signal and the length of defect free sub-sequences suggests that defects mainly affect the zippering of the DNA strands. Zeng and Zocchi \cite{Zeng2006} observed an essentially two-state melting process for short single MM oligonucleotide duplexes. They propose two mechanisms how mismatch defects could affect duplex stability. 1. Defects may lower the barrier for bubble  formation. 2. A defect could introduce 'extra ends' in the middle of the duplex, thus resulting in two separate, effectively shortened duplexes, with reduced stabilities.
\newline
It is uncertain whether defects can lower the threshold for bubble formation so much that denaturation bubbles in short duplexes (in our experiments 25 bases) become relevant in comparison to partial denaturation from the end. The very similar impact of various mismatch and bulge defects on the HS also doesn't favor the idea of bubble formation, as distinct differences in the impacts of the defect types  might be expected. Also, the surprisingly large HS of Group II single bulges, which from a static point of view do not differ from Group I bulges, could be better explained with a kinetic zipper model of the DNA duplex (the latter effect is discussed further below).\newline
We suggest that defects may act as kinetic barriers, preventing the rapid zipping of the strands. In PM duplexes, at temperatures below the PM melting temperature, complete unzipping of the duplex can be regarded as a very rare stochastic event, since the zipping rate $k_{+}$ is  4 to 9 times larger \cite{Craig1971} than the unzipping rate $k_{-}$. 
A defect, due to its unsuitable conformation, may significantly reduce the zipping rate $k_{+}$ at the defect site. Since the duplex is kinetically trapped in a weakly hybridized state with only few remaining base pairs, the probability for a complete dissociation, and thus the dissociation rate, is increased in comparison to the PM duplex.\newline
The destabilizing effect is expected to be position-dependent. Its impact is largest if the defect is located in the middle of the duplex, since then, in the kinetically trapped conformation, the duplex is unzipped by more than 50\%. Depending on the position of the defect the dynamic equilibrium of continuous partial melting and renaturation is shifted towards that of shorter effective duplex lengths, with an increased probability for complete strand dissociation. 
\newline
Sterical crowding at the surface \cite{Peterson2002} could possibly introduce a positional dependence on the HS of defect probes. Reduced accessibility of the probes surface-bound 3'-ends can in principle decrease the impact of defects located near the 3'-end, and thus result in increased hybridization signals of the corresponding probes. This runs contrary to the largely symmetrical intensity profiles observed (cf. Figure \ref{fig:COM5MMProfileFig3}) and does therefore not provide an explanation for the impact of defect position. 
Pozithkov \textit{et al.} \cite{Pozhitkov2006} report increased HS from probes with MM defects near the surface-bound 3'-end. This bias is in accordance with results from our two-deletion-experiment (Figure \ref{fig:TwoDelHSFig11}). In our case the bias could also originate from a rather limited set of  9 probe sequence motifs. In other experiments, focussing on single defects, we didn't notice systematically increased HS for 3'-located defects.
\subsection*{MM type and nearest neighbor influence}
We observed that single-base MMs introduced at the site of a C$\cdot$G base pair result in a larger decrease of the hybridization signal (in respect to the PM hybridization signal) than MM defects affecting  A$\cdot$T base pairs. The same effect is seen for single base deletions. This indicates that the individual C$\cdot$G base pairs are significantly more important for duplex stability than A$\cdot$T base pairs. 
The effect seems to largely determine the impact of the different MM types  X$\cdot$Y on the HS.
Our experimental results, in accordance with nearest-neighbor thermodynamic parameters for Watson-Crick base pairs \cite {SantaLucia2004}, reflect the increased base stacking and hydrogen bonding interactions of C$\cdot$G base pairs.
\newline
The type-specific impact of MMs is shown in the following series (see also Figure \ref{fig:MMTypeBoxPlotFig5}), which is ordered according to median hybridization signals obtained from 16mer single base MM duplexes (DNA/DNA). The largest hybridization signal, corresponding to the least impact, is observed for G$\cdot$A.
\newline G$\cdot$A$>$T$\cdot$T$\ge$G$\cdot$T$>$A$\cdot$G$\ge$C$\cdot$T$\approx$C$\cdot$A$>$A$\cdot$A$\approx$T$\cdot$G$\approx$C$\cdot$C$\approx$T$\cdot$C $\ge$A$\cdot$C$\ge$G$\cdot$G. This order is in good agreement with a similar order of normalized HS determined by Wick \textit{et al.} \cite{Wick2006}. A significant difference is the mismatch base pair G$\cdot$G which Wick and also Sugimoto \textit{et al.} \cite{Sugimoto2000} have identified as a relatively stable MM base pair. Interestingly, in accordance with our results, Pozithkov \textit{et al.} \cite{Pozhitkov2006} have recently reported  G$\cdot$G to be among the least stable mismatch base pairs.
\newline
We find that (with the exception of A$\cdot$G) mismatch types affecting C$\cdot$G base pairs result in lower HS than those affecting A$\cdot$T base pairs. The difference between mismatch types X$\cdot$Y and Y$\cdot$X  (a similar observation with RNA/DNA duplexes has been described in \cite{Pozhitkov2006}) is introduced by the comparison (due to normalization) with the corresponding PM hybridization signal. Here we find again that defects affecting C$\cdot$G base pairs result in a larger decrease of the HS (in respect to the PM hybridization signal) than defects affecting A$\cdot$T base pairs.
\newline
Categorization of mismatch types into two groups affecting either A$\cdot$T or C$\cdot$G base pairs (this also separates the  MM types X$\cdot$Y and Y$\cdot$X) results in two equivalent orders of MM affinities:
GA$>$GT$>$CT$\approx$CA and AG$>$TG$\ge$TC$\ge$AC
\newline
The variation of the HS observed for the particular MM types is largely due to nearest neighbor effects. C$\cdot$G base pairs next to the MM defects  (for most MM types) result in significantly increased HS. 
This 'closing base pair effect' has also been described in \cite {SantaLucia2004}.

\subsection*{Stabilization of Group II single base bulges}
We observe  significantly increased HS  of single-base insertion defects in which the insertion base is placed next to a like-base. Increased stability of Group II bulges in comparison with Group I bulges is well known \cite{Ke1995,Zhu1999,Znosko2002}. According to Ke and Wartell \cite{Ke1995} the increased stability of Group II bulges originates from positional degeneracy of the extra unpaired base. Additional conformational freedom, entailaing higher entropy, results in lowered duplex free energy. 
According to  Zhu \textit{et al.} \cite{Zhu1999} position degeneracy accounts for an average stabilization of -0.3 to -0.4 kcal/mol (in good agreement with the theoretical estimate of -RT ln 2 = -0.43 kcal/mol at 37\degreesC) for a two position degeneracy.  Znosko \textit{et al.} \cite {Znosko2002} reported that Group II duplexes are on average $\delta\Delta G_{37}$=-0.8 kcal/mol more stable than Group I duplexes.
We find that the insertion of G next to T often results in relatively high duplex stability, comparable to a Group II insertion. This observation correlates well with the relatively high stability observed for G$\cdot$A mismatches, suggesting that the  G$\cdot$A mismatch base pair could be degenerate with the neighboring A$\cdot$T base pairs.
\newline
For explanation of the high stability of Group II  duplexes we propose the following mechanism based on a molecular zipper model of the duplexes: 
After nucleation or partial unzipping of the duplex, extending beyond the defect site,
the surplus base might act as barrier, interrupting the rapid zipping (renaturation) of the duplex. Frameshift due to the additional base doesn't allow hybridization beyond that point, resulting - depending on the defect position - in a weakly bound partially zipped duplex. Duplex closure can  only progress if the surplus base is making room (forming a looped out or stacked conformation), thus allowing the subsequent base to form a Watson-Crick base pair with the complementary base in the target strand. From this point zipping can continue rapidly.
\newline
For Group II sequences there is an important difference: The zipping is interrupted at the defect site, which is located at the end of the group of degenerate (identical) bases. As with Group I sequences, at the barrier partial unzipping is likely to occur. There is now an increased probability that one of the degenerate bases makes way (i.e. goes into bulge conformation) and allows the subsequent base to form a base pair. Then, as the frameshift is now compensated, the rapid zipping, resulting in a stabilized duplex, can continue.
Duplex formation for Group I bulges is expected to be slower because for Group I bulges it is necessary that the particular (non-degenerate) surplus base gets into the suitable conformation.
With Group II insertions the barrier, which is trapping the duplex in a weakly bound partially zipped state, can be overcome faster, resulting in  increased stability.



\clearpage


\end{document}